\documentclass[10pt,twocolumn,twoside]{IEEEtran}

\usepackage{xcolor}
\usepackage{cite}
\usepackage{amsmath}
\usepackage{amsfonts} 
\usepackage{amssymb}
\usepackage{tabularx}
\usepackage[utf8]{inputenc}
\usepackage{tikz}
\usepackage{standalone}
\usepackage{bbm}
\usepackage{multirow}
\usepackage{subcaption}
\usepackage{cite}
\usepackage{makecell}

\usetikzlibrary{arrows}
\usetikzlibrary{decorations.pathmorphing}
\usetikzlibrary{shapes}
\usetikzlibrary{backgrounds}
\usetikzlibrary{positioning}
\usetikzlibrary{fit}
\usetikzlibrary{petri}
\usetikzlibrary{patterns}
\usetikzlibrary{matrix}
\usetikzlibrary{automata}
\usetikzlibrary{decorations.pathreplacing}
\usetikzlibrary{angles}
\usetikzlibrary{quotes}

\newtheorem{theorem}{\bf Theorem}
\newtheorem{lemma}[theorem]{\bf Lemma}

\newtheorem{example}{\bf Example}
\newtheorem{remark}{\bf Remark}
\newtheorem{proof}{\bf Remark}

\renewcommand{\forall}{\text{ for all }}
\newcommand{\ex}[1]{\mathbb{E}\left[#1\right]}

\newcommand{\trace}[1]{\text{tr}\left(#1\right)}
\newcommand{\tx}{\textsf{Tx}}
\newcommand{\indep}{\perp \!\!\! \perp}
\newcommand{\pone}{P^{(1)}}
\newcommand{\ptwo}{P^{(2)}}
\newcommand{\specialcell}[1]{\ifmeasuring@#1\else\omit$\displaystyle #1$\ignorespaces\fi}
\def\QED{~\rule[-1pt]{5pt}{5pt}\par\medskip}
\renewenvironment{proof}{{\bf Proof: \ }}{ \hfill \QED}

\begin{document}
\title{Nash Equilibrium Control Policy against Bus-off Attacks in CAN Networks}
\author{Jiacheng~Tang,
        Shiping~Shao,
        Jiguo~Song,
        Abhishek~Gupta
\thanks{Jiacheng Tang, Shiping Shao, and Abhishek Gupta are with the Department of Electrical and Computer Engineering at The Ohio State University, Columbus, OH, USA. Email: {\tt\small tang.481@osu.edu, shao.367@osu.edu, gupta.706@osu.edu}.}
\thanks{Jiguo Song is with Ford Motor Company, Dearborn, MI, USA. Email: {\tt\small jsong26@ford.com}.}
\thanks{The research was supported through a University Alliance Project from the Ford Motor Company.}}
\maketitle
\thispagestyle{empty}
\pagestyle{empty}

\begin{abstract} 
    A bus-off attack is a denial-of-service (DoS) attack which exploits error handling in the controller area network (CAN) to induce an honest node to disconnect itself from the CAN bus. This paper develops a stochastic transmission policy as a countermeasure for the controller-transmitter pair against the bus-off attack. We model this as a non-zero-sum linear-quadratic-Gaussian game between the controller-transmitter pair and the attacker. We derive Nash equilibria of the game for two different information structures of the attacker. We show that the attacker has a dominant attack strategy under both information structures. Under the dominant attack strategy, we show that the optimal control policy is linear in the system state. We further identify a necessary and a sufficient conditions on the transmission policy to have bounded average cost. The theoretical results are complemented by a detailed case study of a bus-off attack on a vehicular adaptive cruise control model.
\end{abstract}


\section{Introduction}
Bus-off attack leverages the standard error handling method of several commonly used in-vehicle networks. Using classic controller area network (CAN) as an example, we review some basics of the CAN data frame that is related to the bus-off attack. In a classic CAN data frame, an 11-bit long field at the beginning of each frame is called an identifier. Each electronic control unit (ECU) attached to the CAN network can define a set of CAN data frames for transmitting and receiving, and each CAN data frame is assigned a unique identifier, where smaller value of the identifier represents higher priority in the CAN network. Due to the broadcast nature of CAN, two messages are not allowed to be sent simultaneously on the CAN bus. If two ECUs simultaneously attempt to send messages over the CAN bus, then the message with the smaller identifier wins the arbitration and gets transmitted first. For each data frame, the actual data field can be encoded in up to 8 bytes. The coding book for the data field, sometimes referred to as a \texttt{.dbc} file, is often different across vehicle's years, models, and brands. Also most such coding books are OEM specific and proprietary. However, by physically attaching the CAN bus and monitoring the traffic, certain part of the coding book of interest can be deduced for some specific vehicles via reverser engineering.

Each ECU is equipped with an error counter to handle errors in messages sent on the CAN bus -- if an ECU sends a message to the CAN bus and detects a conflict, meaning the
CAN message won the arbitration with a smaller value of the identifier but the data field is incorrect, e.g., bus fault or CRC check failure, then it will cause an increment in the error counter. On the other hand, if the data field is correct, then the error counter will decrease with a saturation at 0. If the error counter exceeds a pre-defined threshold, the ECU will switch to a bus-off mode and no further messages to or from this ECU will be sent on the CAN bus, until the ECU is reset or power cycled.

Intelligent attackers can use this error handling feature in the CAN protocol to eventuate an ECU into the bus-off mode. In this case, a compromised ECU (the attacker) in the CAN network could send a message with the exact same identifier as the targeted healthy ECU with arbitrary data field to trigger conflicts. This attack is called bus-off attack -- it requires only the knowledge of the identifier used by the target ECU without any reverse engineering of the encoded data field. Once there are sufficient conflicts within a certain time period, the attacker could then trigger the bus-off event and eventually disconnect the target ECU from the CAN network.

\subsection{Related Works}
Bus-off mechanism is designed to be an error confinement mechanism for CAN network since 1990s \cite{pazul1999controller}. In 2016, the bus-off attack threat was investigated by Cho and Shin \cite{cho2016error}. In 2018, Iehira et al. \cite{iehira2018spoofing} leveraged bus-off attack in the lab environment to completely prevent the transmission of regular messages sent from a target electronic control unit (ECU) even if the target ECU itself is not compromised. Around the same time, Souma et al. \cite{souma2018counter} introduced counter attack as a potential countermeasure such that the node initiated bus-off attack itself falls into bus-off mode before the target node does. Later, this counter attack strategy was improved by Takada er al. \cite{takada2019counter} that enhances the success rate of putting attacker into bus-off mode. In 2019, Ning et al. \cite{ning2019attacker} proposed local outlier factor (LOF) as an intrusion detection algorithm to detect the bus-off attack. Testing the algorithm on a real vehicle shows sufficiently high detection rate and low false alarm rate. In 2021, implementation of a refined bus-off attack strategy on real vehicles called WeepingCAN \cite{bloom2021weepingcan} shows stealthiness of bus-off attack in terms of bypassing detection with high success rate. 

To the best of our knowledge, most of the existing works in bus-off attack have been conducted toward demonstrating the attacker's capability, or detecting bus-off attack and assuming attacker node can be removed completely from the network once detected. The moment the attacker starts to transmit messages, some the regular messages that are sent by the target node cannot be successfully delivered. Such regular packet loss can even happens before the bus-off attack succeeds, e.g., before reaching the predefined error counter
threshold. The packet loss due to attack can significantly degrade the performance of the overall control system. The resulting impact on the performance of the control tasks associated with the bus-off attack has not received enough attention. This is the topic we investigate in this paper.

\subsection{Contributions of this paper}
In this paper, we present a mathematical model for the bus-off attack and formulate it as a non-zero sum game between the target node and the attacker. 
A stochastic transmission policy is proposed as a proactive countermeasure when the attacker persists in the network. 
We determine the Nash equilibria of the non-zero sum game in the cases off open-loop and closed-loop attacker. To demonstrate the effectiveness of our theoretical results, we apply the stochastic transmission policy on an adaptive cruise control (ACC) and show that under the Nash equilibrium strategies, an appropriate stochastic transmission policy stabilizes the error counter and the ego vehicle maintains a safe distance with the leading vehicle.


\subsection{Outline}
The paper is organized as follows: The non-zero sum game between attacker and defender is formulated in Section \ref{sec_pf}. Stochastic transmission as the defense policy and some preliminaries on the attack policy is defined in Subsections \ref{sec_txController} and \ref{sec_txAttacker}. In Subsection \ref{sec_MCerrorCounter}, we define a Markov chain model for the error counter and define the bus-off event. The dominant attack policy is derived in Section \ref{sec_dominantAttackProof}. Given the dominant attack strategy, the optimal control under finite and infinite horizon cases are discussed in Section \ref{sec_OptimalControlFinite} and \ref{sec_OptCtrl} respectively. In Section \ref{sec_simulation}, the efficacy of applying stochastic transmission against bus-off attack is evaluated for adaptive cruise control with emergency braking. In Section \ref{sec_conclusion}, we conclude the discussion and present our thoughts on the potential directions for the future work.

\section{Problem Formulation}
\label{sec_pf}
In this section, we formulate the bus-off attack problem in which a controller-transmitter pair is acting against a common adversary that sends messages on the same bus leading to repeated collisions and an eventual bus-off event. To simplify the analysis, we consider the control system model to be discrete linear time-invariant system with a quadratic objective function. As shown in Section \ref{sec_simulation}, a discrete-time Linear-Quadratic-Regulator (LQR) is one of the most commonly used controllers for control over CAN communication, such as longitudinal speed control and lateral steering control.

The time horizon is discrete and takes values in $\mathbb{Z}_{\geq 0}$. Let $x_t\in\mathcal X$ denote the state of the system, $u_t\in\mathcal{U}$ denote the controller's action, and $v_t$ denote the actuation noise at time $t$. We use $\alpha_t\in\{0,1\}$ to denote the transmitter's action, in which $\alpha_t = 1$ means that the transmitter decides to transmit the control signal to the actuator at time $t$ and $\alpha_t = 0$ means that the transmitter decides not to transmit the control signal. Similarly, $\beta_t\in\{0,1\}$ denotes the transmission action of the attacker at time $t$. In the event of a collision at time $t$, i.e., $\alpha_t = \beta_t = 1$, no actuation signal is received by the controller and zero control is applied to the system. Accordingly, the system model can be written as 

\begin{figure}[t]
    \centering
    \includegraphics[width = 0.48\textwidth]{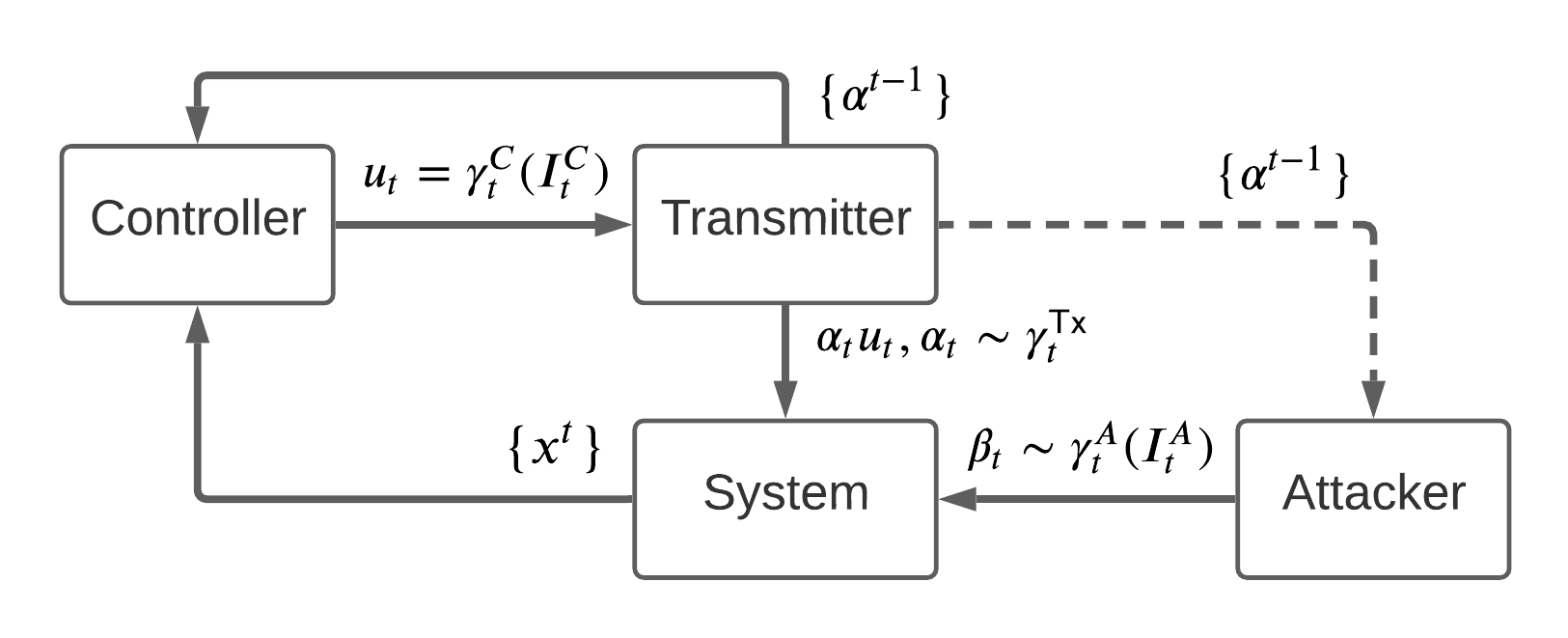}
    \caption{\textbf{Information Flow Diagram:} The history $\{\alpha^{t-1}\}$ of the transmitter's decisions is only available to the closed-loop attacker and is marked as dashed arrows.}
    \label{Fig_Overall}
\end{figure}

\begin{equation}
\label{eqn_systemModel0}
    x_{t+1} = Ax_t + (1-\beta_t)\alpha_tBu_t + v_t.
\end{equation}
Here, $v_t\overset{iid}{\sim}\mathcal{N}(0,\Sigma_v)$ is assumed to be a zero-mean Gaussian random vector with a known covariance matrix. Note that the control action $u_t$ is only applied to the state $x_t$ when the transmitter decides to transmit $\{\alpha_t=1\}$ and the attacker decides not to transmit a message $\{\beta_t=0\}$.

For the transmitter, the decisions $(\alpha_t)_{t\in\mathbb{Z}_{\geq 0}}$ is assumed to be a memoryless stochastic process with Bernoulli distribution  parameterized by transmission probability $p$.

For the controller, let $x^t = (x_0,\ldots,x_t)$ and $\alpha^t = (\alpha_0,\ldots,\alpha_t)$ be the history of state and action respectively. The  information set available to the controller at time $t$ is denoted by $I^C_t = (x^t,\alpha^{t-1})$. Let $\mathcal I^C_t$ denote the set of all possible realizations of the information at the controller at time $t$. We let $\gamma^C_t:\mathcal I^C_t\rightarrow\mathcal{U}$ denote the controller's policy at time $t$. Further, let the control strategy of the controller be the collection of control policies  $\boldsymbol \gamma_C=(\gamma_t^C)_{t\in\mathbb{Z}_{\geq 0}}$ and let $\Gamma_C$ denote the set of all control strategies of the controller. 

For the attacker, we consider two different information structure depending on whether or not it can observe the transmission decisions $\alpha_t$ made by the transmitter:
\begin{itemize}
    \item The space of closed loop attack policy as $\Gamma_A$, with $I^A_t=\{\alpha^{t-1}\}$ containing the history of the transmitter's decisions. The details of the information set and the restrictions on the policy space for this case is explained in Section \ref{sec_txAttacker}.
    \item The space of open loop memoryless attack policy as $\Gamma_A'$, with $I^A_t = \emptyset$ and $\gamma^A_t$ restricted to be a Bernoulli random variable with probability $p'\in [0,p]$.
\end{itemize}
The attacker's policy at time $t$ is denoted by $\gamma^A_t:\mathcal I^A_t\rightarrow\wp(\{0,1\})$. A similar convention is adopted for $\boldsymbol\gamma_A$ and $\Gamma_A$. 

Given a finite horizon $N$, the cost function for controller, $J_C$, is a quadratic function defined as
\begin{equation}
\begin{aligned}
    &J_C^N(\gamma_C,\gamma_A)\\
    &\quad=\mathbb{E}\bigg[x_N^TQx_N+\sum_{t=0}^{N-1} x_t^TQx_t + (1-\beta_t)\alpha_tu_t^TRu_t\bigg],
\end{aligned}
\end{equation}
where $Q\geq0$ and $R>0$. Moving toward the infinite horizon case, we adopt average performance with the same parameters $Q$ and $R$ as the finite case but no terminal cost, that is
\begin{align*}
    &J_C^\infty(\gamma_C,\gamma_A)\\
    &\quad\triangleq\limsup_{N\rightarrow\infty}\frac{1}{N}\mathbb{E}\Bigg[\sum_{t=0}^{N-1} x_t^TQx_t + (1-\beta_t)\alpha_tu_t^TRu_t\Bigg].
\end{align*}

From attacker's perspective, denote $S_t$ as the error counter at time $t$, and define the dynamics of error counter by
\begin{equation*}
    S_t = \left\{\begin{array}{ll}
    \min(\bar e,S_{t-1}+e_+), & \mbox{if } \alpha_t=\beta_t=1 \\ \max(0,S_{t-1}+e_-), & \mbox{if } \alpha_t=1, \beta_t=0 \\
S_{t-1}, & \mbox{otherwise}
\end{array} \right.,
\end{equation*}
with $S_0=0$, where $e_+>0$ is the penalty of collision, $e_-<0$ is the decrements of error counter in the case of a successful transmission, and error counter is bounded below by 0, and above by a threshold $\bar e$, which are all pre-defined constants. The bus-off event is then defined as the stopping time 
$$\xi=\min\{t:S_t\geq \bar e\},$$
which is the first time the error counter exceeds the threshold $\bar e$. The attacker's objective function $J_A$ is the expected number of messages the attacker needs to trigger the bus-off event, that is
\begin{align*}
    J_A(\gamma_C,\gamma_A)=\ex{\xi|\gamma_C,\gamma_A}.
\end{align*}

Now we are interested in computing a subgame-perfect Nash equilibrium $(\gamma_C^*,\gamma_A^*)$ of the nonzero-sum game between the controller and the attacker under the two information structures of the attacker such that
\begin{align*}
    J_C(\gamma_C^*,\gamma_A^*)&\leq J_C(\gamma_C,\gamma_A^*)\,\text{, for all } \gamma_C,\\
    J_A(\gamma_C^*,\gamma_A^*)&\leq J_A(\gamma_C^*,\gamma_A)\,\text{, for all } \gamma_A.
\end{align*}

\subsection{Main Results} \label{sec_pf_mainResults}
We list the main results of this paper and the detailed proofs are presented in the later sections. The first result here shows there is a dominant attack strategy for the case of closed loop attacker.
\begin{theorem} \label{thm_DominantAttack}
There exists a dominant closed-loop policy $\gamma_A^*\in\Gamma_A$ such that for any $\gamma_C\in\Gamma_C$
$$J_A(\gamma_C,\gamma_A^*)\leq J_A(\gamma_C,\gamma_A),\,\forall \gamma_A\in\Gamma_A.$$
Under the dominant attack policy, we have $\beta_t = \gamma_t^{A*}(I^A_t) = \alpha_{t-1}$.
\end{theorem}
\begin{proof}
Please refer to Subsection \ref{sec_dominantAttackProof}.
\end{proof}
With some minor changes in the proof, the second result shows there is also a dominant attack strategy for the case of open loop attacker.
\begin{theorem} \label{thm_nonadaptiveAttackNonadaptive}
There exists a dominant open-loop policy $\gamma_A'^*\in\Gamma_A'$ for the attacker such that 
$$J_A(\gamma_C,\gamma_A'^*)\leq J_A(\gamma_C,\gamma_A'),\,\forall \gamma_A'\in\Gamma_A'.$$
\end{theorem}
\begin{proof}
Please refer to Subsection \ref{sec_nonadaptiveAttack}.
\end{proof}
Under the dominant attack strategy, the following result shows the optimal control strategy for the case of closed loop attacker.
\begin{theorem}\label{thm_OptimalControlFinite}
There exists an optimal closed-loop control policy $\gamma_C^*\in\Gamma_C$ such that
$$J_C(\gamma_C^*,\gamma_A^*)\leq J_C(\gamma_C,\gamma_A^*),\,\forall \gamma_C\in\Gamma_C,$$
where $\gamma_A^*$ is derived in Theorem \ref{thm_DominantAttack}. Here, $\gamma_i^{C^*}$ is linear in $x_i$, for all $i\in\{0,...,N-1\}$. Further, in the infinite horizon case, there exists a $\rho_{\min}$ such that if 1) system parameter $(A,B)$ and $(A,Q^{\frac{1}{2}})$ are controllable, and 2) the given transmission policy with $p$ satisfies $p(1-p)>\rho_{\min}$, then
$$J_C^\infty(\gamma_C^*,\gamma_A^*)<\infty.$$
\end{theorem}
\begin{proof}
Please refer to Section \ref{sec_OptimalControlFinite} for the finite horizon cost $J_C$, and Section \ref{sec_OptCtrl} for the infinite horizon average cost $J_C^\infty$ as $N\rightarrow\infty$.
\end{proof}
The linearity of optimal control and the condition for bounded average cost as $N\rightarrow\infty$ for the open-loop attack case can be proved in a similar way, and is  discussed in Section \ref{sec_OptCtrl}. 

We next add some more details of system with attacker and defender in the following section. This serves as the preliminaries of computing the subgame-perfect Nash equilibrium strategies of the players.

\section{System Modeling under the Bus-off Attack} \label{sec_pre}
\subsection{Transmission policy} \label{sec_txController}
Recall that the transmission decisions over time space is represented by $(\alpha_t)_t$ in the system model \eqref{eqn_systemModel0}, which determines a counting process for the transmission of the messages. In the message space, let $t_i^\tx$ denote time elapsed between the $(i-1)^{\text{th}}$ and $i^{\text{th}}$ message transmissions. In this paper, we further restrict the distribution of $(\alpha_t)_t$ to i.i.d. Bernoulli distribution with parameter $p$. Thus, $(t_i^\tx)_i$ is a geometrically distributed sequence of random variables, and we have
$$[\alpha_t|\tilde{I}_t]\overset{d}{=}\alpha_t\,\forall t \iff t_i^\tx\overset{iid}{\sim} \text{Geometric}(p), p\in(0,1).$$

\begin{example}
The sequence of realizations $(t_i^\tx)_{i=1}^3=(1,3,2)$ over message space uniquely determines $(\alpha_t)_{t=1}^6=(1,0,0,1,0,1)$ over time space.
\end{example}


\subsection{Attack Policy} \label{sec_txAttacker}
In this section, we will discuss two types of attack policy, where the closed loop policy requires the history of the transmission policy, and the open loop policy requires no such information. We constrain the attacker to drop messages at the same frequency (on average) as the transmitter. Otherwise, the attacker will just choose to attack the network every single time, in which case the control signal $\{u_t\}_t$ will never be successfully delivered.

\subsubsection{Closed loop Attack Policy}
Let $t_i^A$ be the time elapsed between $(i-1)^{\text{th}}$ message from transmitter and the $i^{\text{th}}$ blocking attempt from the attacker. The following three situations arise:
\begin{itemize}
    \item If $t_i^A<t_i^\tx$, then the attacker will send a message prior to the $i^{\text{th}}$ message from the transmitter. No collision will happen at the time of sending $i^{\text{th}}$ message from transmitter, hence the attack is launched and failed to block the message and increase the error counter.
    \item If $t_i^A=t_i^\tx$, then the $i^{\text{th}}$ message sent from transmitter triggers a collision, and the attack is launched and is successful in increasing the error counter.
    \item If $t_i^A>t_i^\tx$, then before the attacker decided to cause a collision, the $i^{\text{th}}$ message from the transmitter will be sent. In this case, attacker's waiting time for the next blocking attempt will be reset at the time of observing $i^\text{th}$ message sent from the transmitter. In this case, the $i^\text{th}$ attempt of the attacker is withdrawn.
\end{itemize}


Now we denote closed loop attack policy $\gamma_A\in\Gamma_A$ as any distribution supported by $\mathbb{Z}_{>0}$, where $t_i^A\overset{iid}{\sim}\gamma_A$. Note that $\gamma_A$ can be parameterized as $\{\iota_k\}_{k\in\mathbb{Z}_{>0}}$, where
$$\mathbb{P}(t_i^A=k)=\iota_k, \forall i,k\in\mathbb{Z}_{>0}.$$
Since $\{t_i^A\}_i$ are independent, in the time space we have
$$[\beta_t|I^A_t]=[\beta_t|\alpha_{1:t-1},\beta_{1:t-1}]\overset{d}{=}[\beta_t|\alpha_{k_t:t-1},\beta_{k_t:t-1}].$$
where $k_t=\max\{k<t:\alpha_k=1\}$. This implies the distribution of $\beta_t$ only depends on the history of up to $k_t$, where $k_t$ is the last time step when the transmitter sends a packet.

\subsubsection{Open loop Attack Policy} The open loop policy considers the case when the attacker requires no information to make the attack decision $\beta_t$. Similar to the previous section, this implies
$$t_i^A\overset{iid}{\sim}\text{Geometric}(p'),\quad \beta_t\overset{iid}{\sim}\text{Bernoulli}(p').$$
As a result, the attack policy in the reduced space $\Gamma_A'$ of open loop attack can be parameterized by $p'$, and we write
$$\gamma_A'=p'.$$
Under the constraint when the attacker could have at most the same transmission frequency (on average) as the transmitter, we have $p'\leq p$. 

\begin{figure*}[!t]
\centering
\begin{tikzpicture}[>=stealth', shorten >=1pt, auto,
    node distance=2.5cm, scale=1, 
    transform shape, align=center, 
    state/.style={circle, draw, minimum size=1.2cm},font = \LARGE]

\node[state] at (0,0) (0) {$0$};
\node[state] at (3,0) (1) {$1$};
\node at (6,0) (1n) {$\cdots$};
\node[state] at (9,0) (1e) {$e_+$};

\node at (12,0) (1n1) {$\cdots$};

\node[state] at (15,0) (2e) {$2e_+$};

\node at (18,0) (2n2) {$\cdots$};
\node at (21,0) (22n2) {$\cdots$};

\node at (24,0) (k2n2) {$\cdots$};
\node at (27,0) (1ke) {$\cdots$};
\node[state] at (30,0) (ke) {$\bar{k}e_+$};

\node at (33,0) (ken) {$\cdots$};
\node[state] at (36,0) (be) {$\bar e$};



\draw[every loop]
(0) edge[bend left, auto = left] node {$q$} (1e)
(1) edge[bend left, auto = left] node {$1-q$} (0)
(0) edge[loop above] node {$1-q$} (0)
(1n) edge[bend left, auto = left]  node {$1-q$} (1)
(1e) edge[bend left, auto = left] node {$1-q$} (1n)
(1n1) edge[bend left, auto = left] node {$1-q$} (1e)
(1) edge[bend left, auto = left] node {$q$} (1n1)

(1e) edge[bend left, auto = left] node {$q$} (2e)
(2e) edge[bend left, auto = left] node {$1-q$} (1n1)

(2n2) edge[bend left, auto = left] node {$1-q$} (2e)
(2e) edge[bend left, auto = left] node {$q$} (22n2)
(22n2) edge[bend left, auto = left] node {$1-q$} (2n2)

(k2n2) edge[bend left, auto = left] node {$q$} (ke)
(ke) edge[bend left, auto = left] node {$1-q$} (1ke)
(1ke) edge[bend left, auto = left] node {$1-q$} (k2n2)

(ken) edge[bend left, auto = left] node {$1-q$} (ke)
(ke) edge[bend left, auto = left] node {$q$} (be)

(be) edge[loop below] node {$1$} (be);

\end{tikzpicture}
\caption{The error counter process $\{S_i\}_i$ forms a Markov chain. With probability $q$, the error counter is increased by $e_+$, and with probability $1-q$ the error counter is decreased by $e_-$. The upper and lower bound of the error counter is given by $\bar e$ and $0$. One can consider $\bar e$ as an absorbing state since we are interested in the time of the error counter cross the thresholds $\bar e$. Here, $\bar{k}\in\mathbb{Z}_{\geq 1}$ is such that $\bar{k}e_+\leq \bar e< (k+1)e_+$.}
\label{fig:mc}
\end{figure*}
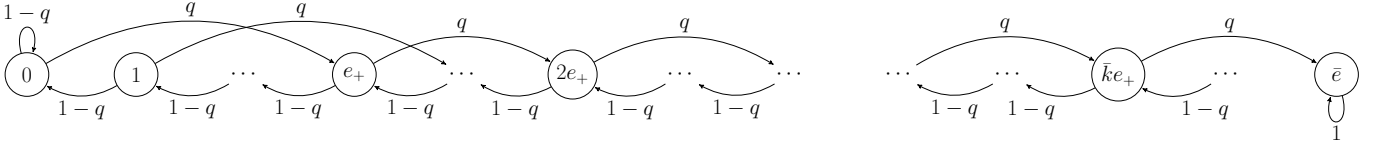

\subsection{Markov Chain Model for Error Counter and Bus-off Event} \label{sec_MCerrorCounter}
Leveraging the transmission policy and the attack policy introduced in the previous two sections, we can derive the dynamics of the error counter based on the collision probability $q$, under the message space. Recall that the penalty of collision is $e_+$ and the reward for successful transmission is $e_-$. Here, we assume these are given constants, with $e_+>0>e_-$ and $\vert e_+\vert>\vert e_-\vert$. Without loose of generality, we set $e_-=-1$, and $e_+,\Bar{e}\in\mathbb{Z}_{>1}$.




If we denote the change of error counter by sending $i^{\text{th}}$ message from the controller by $e_i$, then $e_i$ follows from
\begin{equation*}
      \mathbb{P}(e_i=a)=\left\{
      \begin{array}{@{}ll@{}}
        \mathbb{P}(t_i^\tx=t_i^A) & \text{if}\ a = e_+, \\
        \mathbb{P}(t_i^\tx\neq t_i^A) & \text{if}\ a = e_-, \\
        0 & \text{otherwise,}
      \end{array}\right.
\end{equation*}
where the probability of collision $q\triangleq\mathbb{P}(t_i^\tx=t_i^A)$. For implementation purpose, we consider $S_i$, which is defined as the value of the error counter after sending $i$ messages, to be lower bounded by $0$. Then $S_i$ can be updated by
\begin{align*}
    S_{i+1}=\min\left\{\max\left\{S_i+e_{i+1}, 0\right\},\bar e\right\}.
\end{align*}

Thus, since $e_{i+1}$ is independent of $\{S_i,..,S_1\}$ for any $i\geq 1$, we have $[S_{i+1}|S_i,...,S_1]\overset{d}{=}[S_{i+1}|S_i]$,
which shows that the process $\{S_i\}_{i\in\mathbb{Z}\geq 0}$ is Markovian, which is illustrated in Figure \ref{fig:mc}. Note that the state space of this process is finite, 
$$S_i\in \{[0,\bar e]\cap\mathbb{Z}_{\geq 0}\}=:\mathcal{S}$$
for all $n$ since $\bar e<\infty$, $e_+\in\mathbb{Z}_{>1}$ and $e_-=-1$. 

By having a lower bound of zero on the error counter, it is straightforward to prove that the error counter process shown in Figure \ref{fig:mc} has a single recurrent state, which is the threshold $\bar e$, and all other states are transient. This implies that the bus-off event will occur eventually if the attacker could persist in the network for a sufficiently long time. However, by picking an appropriate transmission probability $p$, a negative drift of the error counter can make the active control period large, that is, the stopping time $\xi$ could be very large with high probability. 

More precisely, the dynamics shown in Figure \ref{fig:mc} can be treated as a discrete-time first-come-first-service G/D/1 queue: the service time is constant with $s(t)=1$, and the arrival process is independent over time with
$$\mathbb{P}(a(t)=e_+-1)=q,\quad\mathbb{P}(a(t)=0)=1-q.$$ In \cite{chang1995sample} (Section 4), the results there proved that the tail probability of the queue length seen by a typical customer decreases exponentially with respect to the threshold $\bar e$, when $\bar e$ is large.

\section{Attacker's dominant strategy} 
\label{sec_proofDominantAttack}
In this section, we will prove the close-loop attacker's dominant strategy stated in Theorem \ref{thm_DominantAttack}, with the open-loop case following similar lines of arguments. Given the dominant attack strategy in each cases, we will then derive the induced state space model based on the general model defined in \eqref{eqn_systemModel0}. The induced models will be used later in Section \ref{sec_OptimalControlFinite} to further analyze the optimal control strategy.

\subsection{Dominant Closed-loop Attack Strategy} \label{sec_dominantAttackProof}
It is clear that transmission probability $p$ does not depend on $\gamma_C$, and therefore, $J_A$ is independent of $\gamma_C$. However, we show in this section that since the transmission policy is restricted to be geometrically distributed, the attacker has a dominant strategy $\gamma_A^*$, which is to jam the channel immediately after a successful transmission from the controller. Recall the probability for the attacker to  jam $k$ steps after each transmission as $\mathbb{P}(t_i^A=k)=\iota_k$ for all $i$, and $\sum_k\iota_k=1$. Let
\begin{align} \label{eq:gammaA}
    \gamma_A^*=(\iota_1=1,\iota_2=0,\iota_3=0...),
\end{align}
then $\gamma_A^*$ is the dominant strategy of attacker, i.e., for all $\gamma_A$, $p\in(0,1)$, and $\gamma_C$,
\begin{align*}
    J_A(\gamma_C,\gamma_A^*)&\leq J_A(\gamma_C,\gamma_A).
\end{align*}
This is proved by the following result.
\begin{theorem}
    For any memoryless transmission policy parameterized by $p\in(0,1)$, we have
\begin{align*}
    \mathbb{E}[\xi|\gamma_A^*]\leq \mathbb{E}[\xi|\gamma_A], \forall \gamma_A\in\Gamma_A.
\end{align*}
where $\gamma_A^*$ defined in \eqref{eq:gammaA} is the unique minimizer.
\end{theorem}
\begin{proof}
    Please refer to Appendix \ref{app_dominantAttack}.
\end{proof}


With dominant attack policy where $t^A_i=1$, we notice that as long as there are two consecutive transmission decisions for the controller to transmit the control action, the second message will be lost in transmission due to collision caused by the attacker's action. The state equation in \eqref{eqn_systemModel0} for case I and II now changes to
\begin{equation}
\label{eqn_statespaceWattack}
    x_{t+1} = Ax_t + (1-\alpha_{t-1})\alpha_tBu_t + v_t,
\end{equation}
with $\alpha_t\overset{iid}{\sim} \text{Bernoulli}(p)$. Thus, the state for the controller in this problem is $[x_t^T,\alpha_{t-1}]^T$. 

\subsection{Dominant Open-loop Attack Strategy} \label{sec_nonadaptiveAttack}
Now consider the case of open loop attack strategies where $\gamma_A'\in\Gamma_A'$, and recall that any $\gamma_A'$ can be parameterized by $p'\in[0,p]$ for a fixed transmission policy with $p$. The dominant open-loop attack strategy is $p'^*=p$, which can be proved by the following result.
\begin{lemma}
For any memoryless transmission policy parameterized by $p\in(0,1)$, we have
\begin{align*}
    \mathbb{E}[\xi|p'^*]\leq \mathbb{E}[\xi|p'], \forall \gamma_A'\in\Gamma_A'.
\end{align*}
\end{lemma}
\begin{proof}
Notice that the attacker's decision is independent of the transmitter's decision, therefore the probability of collision is $q(p')=pp'$. The monotonicity result of $v_0(q)$ proved in Appendix \ref{app_dominantAttack} implies $p'=p$ is the unique optimal strategy.
\end{proof}
Given the dominant attack strategy in the open loop case, we have
$$\alpha_t\overset{iid}{\sim}\text{Bernoulli}(p),\quad\beta_t\overset{d}{=}\alpha_t,\quad\{\alpha_t\}_t\indep\{\beta_t\}_t.$$
As a result, the state space equation in \eqref{eqn_systemModel0} for Case III and IV now reduces to
\begin{equation}
\label{eqn_statespaceNonadaptive}
    x_{t+1} = Ax_t + \tilde{\alpha}_tBu_t + v_t,
\end{equation}
with $\tilde{\alpha}_t\overset{iid}{\sim} \text{Bernoulli}(p(1-p))$, $p\in(0,1)$.

\section{Optimal Control under Finite Horizon} \label{sec_OptimalControlFinite}
In this section, we derive the optimal control strategy of the controller against the dominant attack strategy of the attacker. We consider two cases separately. We first derive the optimal control strategy with closed loop attacker and then proceed to deriving the optimal control strategy with open loop attacker. The optimal controller for the finite horizon would be optimal control strategy when the horizon length $N$ is smaller than the stopping time $\xi$ almost surely. It is easy to see that if $S_0 = 0$, then $\xi\geq \bar e/e_+$ almost surely.

\subsection{Optimal Control with Closed-loop Attacker}
We analyze the case with closed loop attacker with the system model defined in \eqref{eqn_statespaceWattack}. Under the dominant attack strategy, the controller's objective function reduces to:
\begin{equation}
\label{eqn_costFinite}
    \mathbb{E}\left[x_N^TQ_Nx_N + \sum_{t=0}^{N-1} x_t^TQ_tx_t + (1-\alpha_{t-1})\alpha_tu_t^TR_tu_t\right],
\end{equation}
with the information set that is available to the controller at time $t$ as $I_t =  \{x^t,\alpha^{t-1}\}$. We now derive the optimal control policy $\gamma^*$ using dynamic programming. First, notice that the terminal value function is
$$V_N(x_N,\alpha_{N-1}) \triangleq \,\mathbb{E}[x_N^TQx_N|I_N]=x_N^TQx_N.$$
With $I_{N-1}=\{x^{N-1},\alpha^{N-2}\}$, we have
\begin{align*}
    V&_{N-1}(x_{N-1},\alpha_{N-2})\triangleq\underset{u_{N-1}\in\mathbb{R}^m}{\min}\mathbb{E}[x_{N-1}^TQx_{N-1}+\\
    &+(1-\alpha_{N-2})\alpha_{N-1}u_{N-1}^TRu_{N-1}+V_N(x_N)\vert I_{N-1}]\\
    =&\underset{u_{N-1}\in\mathbb{R}^m}{\min}x_{N-1}^TQx_{N-1}+p(1-\alpha_{N-2})u_{N-1}^TRu_{N-1}\\
    &+\mathbb{E}[V_N(x_N)\vert I_{N-1}],
\end{align*}
where the second equality holds since $\alpha_{N-2}$ is known given $I_{N-1}$ and $\alpha_{N-1}$ is random given $I_{N-1}$ and is independent of $\alpha_{N-2}$. Expanding the remaining term using the state dynamics, we have
\begin{align*}
    &\mathbb{E}[V_N(x_N,\alpha_{N-1})\vert I_{N-1}]=\mathbb{E}[x_N^TQx_N\vert I_{N-1}]\\
    =&x_{N-1}^TA^TQAx_{N-1}+2p(1-\alpha_{N-2})x_{N-1}^TA^TQBu_{N-1}\\
    &+p(1-\alpha_{N-2})u_{N-1}^TB^TQBu_{N-1}+tr(Q\Sigma_v),
\end{align*}
where $\alpha_i^2\overset{d}{=}\alpha_i$ since $\alpha_i\in\{0,1\}$. This yields the value function at step $N-1$ as
\begin{align*}
    &V_{N-1}(x_{N-1},\alpha_{N-2})\\
    =&\underset{u_{N-1}\in\mathbb{R}^m}{\min}p(1-\alpha_{N-2})u_{N-1}^T(B^TQB+R)u_{N-1}+\\
    &+2p(1-\alpha_{N-2})x_{N-1}^TA^TQBu_{N-1}+\\
    &+x_{N-1}^T(A^TQA+Q)x_{N-1}+tr(Q\Sigma_v),
\end{align*}
which is a quadratic function in $u_{N-1}$. Setting the first derivative with respect to $u_{N-1}$ to 0, we compute the minimizer $u_{N-1}^*$ to get
\begin{align*}
    u_{N-1}^*\triangleq&\gamma_{N-1}^{C^*}(I_{N-1})= K_{N-1}x_{N-1},
\end{align*}
which is linear in state $x_{N-1}$, and the gain matrix is $K_{N-1}:=-(B^TQB+R)^{-1}B^TQA$. The value function is of the form
\begin{align*}
    &V_{N-1}(x_{N-1},\alpha_{N-2})\\
    & =x_{N-1}^T\left[\pone_{N-1}+(1-\alpha_{N-2})\ptwo_{N-1}\right]x_{N-1}+c_{N-1},
\end{align*}
where
\begin{align*}
    \pone_{N-1}=&A^TQA+Q\\
    \ptwo_{N-1}=&-p(B^TQA)^T(B^TQB+R)^{-1}(B^TQA)\\
    c_{N-1}=&\trace{Q\Sigma_v},
\end{align*}
with the following remarks:
\begin{enumerate}
    \item $R$ is a positive definite (PD) matrix by definition, thus $B^TQB+R$ is invertible.
    \item Given symmetric $Q$ and $R$ with $Q$ being PSD and $R$ being PD, $\pone_{N-1}$ and $-\ptwo_{N-1}$ are symmetric and PSD.
\end{enumerate}
We next use induction to prove that given $V_k$, the value function $V_{k-1}$ can be written in the same form.
\begin{theorem} \label{thm_optimalFinite}
Consider the state transition function in \eqref{eqn_statespaceWattack}. Suppose that the value function at time $k$ is given by
\begin{align} \label{eqn_valueFn}
    V_k(x_k,\alpha_{k-1})=&x_k^T\left[\pone_k+(1-\alpha_{k-1})\ptwo_k\right]x_k+c_k.
\end{align}
Then, the value function at time $k-1$ is given by
\begin{align*}
    V_{k-1}(x_{k-1},\alpha_{k-2})=&x_{k-1}^T\left[\pone_{k-1}+(1-\alpha_{k-2})\ptwo_{k-1}\right]x_{k-1}\\
    & +c_{k-1}
\end{align*}
with optimal control $\gamma_{k-1}^{C*}$ given by
\begin{equation*}
    u_{k-1}^*\triangleq\gamma_{k-1}^{C^*}(I_{k-1})=K_{k-1}x_{k-1},
\end{equation*}
where
\begin{align*}
    P_k\triangleq&\pone_{k}+(1-p)\ptwo_{k},\\
    \pone_{k-1}=&A^TP_kA+Q,\\
    \ptwo_{k-1}=&-p(B^T\pone_{k}A)^T(B^T\pone_{k}B+R)^{-1}(B^T\pone_{k}A),\\
    c_{k-1}=&\trace{P_k\Sigma_v}+c_{k},\\
    K_{k-1}=&-(B^T\pone_kB+R)^{-1}B^T\pone_kA.
\end{align*}
\end{theorem}
\begin{proof}
Notice that since $\alpha_{k-2},\alpha_{k-1}\in\{0,1\}$, we have
    \begin{align*}
        &\ex{(1-\alpha_{k-2})^2\alpha_{k-1}^2|I_{k-1}}=p(1-\alpha_{k-2}),\\
        &\ex{(1-\alpha_{k-1})\alpha_{k-1}|I_{k-1}}=0.
    \end{align*}
We then have the following
    \begin{align*}
        &\ex{V_k(x_k,\alpha_{k-1})\vert I_{k-1}}\\
        =&\ex{x_k^T\left[\pone_k+(1-\alpha_{k-1})\ptwo_k\right]x_k+c_k\vert I_{N-1}},\\
        =&x_{k-1}^TA^TP_kAx_{k-1}\\
        &+2p(1-\alpha_{k-2})x_{k-1}^TA^T\pone_{k}Bu_{k-1}\\
        &+p(1-\alpha_{k-2})u_{k-1}^TB^T\pone_{k}Bu_{k-1}+\trace{P_k\Sigma_v}.
    \end{align*}
Now, we apply the dynamic programming step to obtain the value function $V_{k-1}$ as
    \begin{align*}
       & V_{k-1}(x_{k-1},\alpha_{k-2})\\
        =&\underset{u_{k-1}}{\min}p(1-\alpha_{k-2})u_{k-1}^T(B^T\pone_{k}B+R)u_{k-1}\\
        &+2p(1-\alpha_{k-2})x_{k-1}^TA^T\pone_{k}Bu_{k-1}\\
        &+x_{N-1}^T(A^TP_kA+Q)x_{N-1}+\trace{P_k\Sigma_v}+c_k.
    \end{align*}
Following the same argument as in the step $N-1$, this implies
    \begin{align*}
        u_{k-1}^*=&-(B^T\pone_kB+R)^{-1}B^T\pone_kAx_{k-1}\\
        \triangleq&K_{k-1}x_{k-1}.
    \end{align*}
Simple algebraic steps yields the expressions for $\pone_{k-1}$, $\ptwo_{k-1}$ as stated in the statement. This completes the proof.
\end{proof}
     
\subsection{Optimal Control under Open-Loop Attacker}
We now derive the optimal control strategy for the case of open loop attacker using a similar approach as above. The key result is stated below.
\begin{theorem} \label{thm_optimalFiniteopenLoop}
Consider the state transition function in \eqref{eqn_statespaceWattack}, Suppose that the value function at time $k$ is given by
\begin{align*}
    V_k(x_k)=&x_k^TP_kx_k+c_k.
\end{align*}
Then, the value function at time $k-1$ is given by
\begin{align*}
    V_{k-1}(x_{k-1})=&x_{k-1}^TP_{k-1}x_{k-1}+c_{k-1}
\end{align*}
with optimal control $\gamma_{k-1}^{C*}$ given by
\begin{equation*}
    u_{k-1}^*\triangleq\gamma_{k-1}^{C^*}(I_{k-1})=K_{k-1}x_{k-1},
\end{equation*}
where
\begin{align*}
    P_{k-1}\triangleq&A^TP_kA+Q\\
    &-p(1-p)(B^TP_{k}A)^T(B^TP_{k}B+R)^{-1}(B^TP_{k}A),\\
    c_{k-1}=&\trace{P_k\Sigma_v}+c_{k},\\
    K_{k-1}=&-(B^TP_kB+R)^{-1}B^TP_kA.
\end{align*}
\end{theorem}
\begin{proof}
    The proof follows the same arguments as in Theorem \ref{thm_optimalFinite}. Note that the open-loop attack $\beta_k$ at time $k$ does not depend on the previous decision $\alpha_{k-1}$ made by the transmitter. Thus, the value function $V_k$ here is no longer a function of $\alpha_{k-1}$.
\end{proof}
We now consider the infinite horizon average cost problem in the next section.

\section{Optimal Control under Infinite Horizon} \label{sec_OptCtrl}
As discussed in Section \ref{sec_MCerrorCounter}, we now use the result of optimal control under infinite horizon to approximate the scenario when the active control period is large and the probability of bus-off event within that horizon is negligible. Consider the average cost 
\begin{align*}
    J^\infty_C(\gamma_C,\gamma_A^*) = \limsup_{N\to\infty} \frac{1}{N} J^N_C(\gamma_C,\gamma_A^*).
\end{align*}
From Theorem \ref{thm_optimalFinite}, we conclude that
\begin{align*}
     V_{k-1}(x_{k-1},\alpha_{k-2})\leq x_{k-1}^T\left[\pone_{k-1}+\ptwo_{k-1}\right]x_{k-1}+c_{k-1}.
\end{align*}
Note that $\pone_{k-1}$ and $\ptwo_{k-1}$ depends only on $P_k = \pone_k + (1-p) \ptwo_k$. Thus, to show that the long term average cost is bounded, we only need to show that $P_k$ converges as $N\to\infty$. This is established as follows. The update equation of $P_k\mapsto P_{k-1}$ is given by
\begin{align} \label{eqn_MAREreduced}
    g_{\rho}(P)=&A^TPA+Q\\
    &-\rho(B^TPA)^T(B^TPB+R)^{-1}(B^TPA),\nonumber
\end{align}
with $\rho=p(1-p)$. The convergence of such an update scheme has been analyzed in \cite{sinopoli2004kalman}, which we recall below.
\begin{lemma} \label{thm_infHorizonTCP}
    Let $(A,B)$ and $(A,Q^{\frac{1}{2}})$ be controllable, then there exists a critical value $\rho_{\min}$ such that for all $\rho>\rho_{\min}$, there exists a unique positive definite matrix $P_{\infty}= g_\rho(P_{\infty})$. This can be computed as a limit of the forward recursion $P_{j+1} = g_\rho(P_j)$. The critical value $\rho_{\min}$ is computed by solving for the following linear matrix inequality:
    $$\rho_{\min}=\inf\{\rho:\Psi_\rho(Y,Z)>0,\; 0\leq Y\leq I\},$$
    where
    \begin{align*}
        &\Psi_\rho(Y,Z)=\\
        &\begin{bmatrix}
        Y & \sqrt{\rho}(YA+ZB) & \sqrt{1-\rho}YA \\
        \sqrt{\rho}(A^TY+B^TZ^T) & Y & 0\\
        \sqrt{1-\rho}A^TY & 0 & Y
        \end{bmatrix}.
    \end{align*}
\end{lemma}
\begin{proof}
    See Theorem 5 and Corollary 1 in \cite{sinopoli2004kalman}.
\end{proof}
\begin{remark}
It has been shown in \cite[Lemma 5.4]{schenato2007foundations} that
\begin{align}\label{eqn_pBound}
       1-\frac{1}{\max_i\vert\lambda_i^u(A)\vert^2}\leq\rho_{\min} \leq 1-\frac{1}{\prod_i\vert\lambda_i^u(A)\vert^2} ,
    \end{align}
    where $\{\lambda_i^u(A)\}_i$ is the set of unstable eigenvalues of the matrix $A$. The upper and lower bounds are tight. 
\end{remark}

Let us define the following matrices:
\begin{align*}
    \pone_\infty =&A^TP_\infty A+Q,\\
    \ptwo_\infty=&-p(B^T\pone_\infty A)^T(B^T\pone_\infty B+R)^{-1}(B^T\pone_\infty A),\\
    c_\infty=&\trace{P_\infty\Sigma_v}.
\end{align*}

\begin{theorem}
In the case of closed loop attacker, the subgame-perfect Nash equilibrium strategy for the controller and the corresponding average cost is given by
\begin{align*}
    \gamma^{C*}(x) =& -(B^T\pone_\infty B+R)^{-1}B^T\pone_\infty A x,\\
    V(x_k,\alpha_{k-1}) =&x_k^T\left[\pone_\infty+(1-\alpha_{k-1})\ptwo_\infty\right]x_k+c_\infty
\end{align*}
In the case of open loop attacker, the subgame-perfect Nash equilibrium strategy for the controller and the corresponding average cost is given by
\begin{align*}
    \gamma^{C*}(x) =& -(B^T\pone_\infty B+R)^{-1}B^T\pone_\infty A x,\\
    V(x) =&x^TP_\infty x+c_\infty
\end{align*}
\end{theorem}
\begin{proof}
For the fixed (dominant) strategy of the adversary, the corresponding optimization problem for the controller is described in the previous section. We just use the results from Lemma 5.4, Theorem 5.5, and Theorem 5.6 \cite{schenato2007foundations} to determine the stationary strategy for the controller. This immediately leads the subgame perfect Nash equilibrium strategy for the game considered here.
\end{proof}

\section{Application to Adaptive Cruise Control}
\label{sec_simulation}
In this section, we use vehicular adaptive cruise control as an example to demonstrate the control and error performance when applying stochastic transmission. In particular, we adopt the LQR setup used in \cite{li2010model} as the base model. The simulation results shown in this paper are based on a MATLAB Simulink ACC model we developed with stochastic transmission as an additional functional block to the dynamics explained in \cite{li2010model}. 

For adaptive cruise control, the goal is to keep a safe distance between the ego (self) vehicle and the leading vehicle by controlling the desired longitudinal acceleration. Let's denote the desired longitudinal acceleration as $a_\text{des}$ and the actual acceleration as $a_f$. The simplest way to capture the low level vehicle longitudinal dynamics is by a first order transfer function
$$a_f=\frac{K_L}{T_Ls+1}a_\text{des},$$
where $K_L=1$ and $T_L=0.45$ are used in the simulation. The car following system is then constructed as a three dimensional system with state denoted by $x=[\Delta d,\, \Delta v,\, a_f]^T$, where $\Delta d(\Delta v)$ is the relative distance (velocity) between the ego car and the lead car. $\Delta d>0$ implies that the lead car is in front of the ego car, and $\Delta v>0$ means the lead car is faster than the ego car. The desired distance is denoted by $d_\text{des}$ and is defined as
$$d_\text{des}=\tau_h v_f+d_0.$$
where $\tau_h=2.5s$ is used as the nominal time headway, $v_f$ is the velocity of the ego car in $m/s^2$, and $d_0=5m$ is used as the stopping distance. The continuous time version of the system dynamics can be written by
\begin{align*}
    \dot x(t) &= 
    \begin{bmatrix}
    0 & 1 & -\tau_h \\
    0 & 0 & -1 \\
    0 & 0 & -1/T_L
    \end{bmatrix}x(t)+
    \begin{bmatrix}
    0 \\ 0 \\ T_L
    \end{bmatrix}u(t)+
    \begin{bmatrix}
    0 \\ 1 \\ 0
    \end{bmatrix}v(t),
\end{align*}
where $u=a_\text{des}$ and $v=v_p$ is the velocity of the lead vehicle and is treated as a disturbance. The states are assumed to be directly measured by the sensors on the vehicle. In MATLAB simulation, the above system is discretized using $\texttt{c2d}$ function with $100ms$ as the sampling time. Denote the discretized version of the system dynamics under $100ms$ sampling time as
$$x_{k+1}=Ax_k+Bu_k+Gv_k.$$
The quadratic cost is set with $Q=\text{diag}([0.06,\,0.1,\,0.5])$ as a 3-by-3 diagonal matrix for the state $x$ and $R=1$ for the control $u$. 

\begin{figure}[h]
    \centering
    \includegraphics[width=0.4\textwidth]{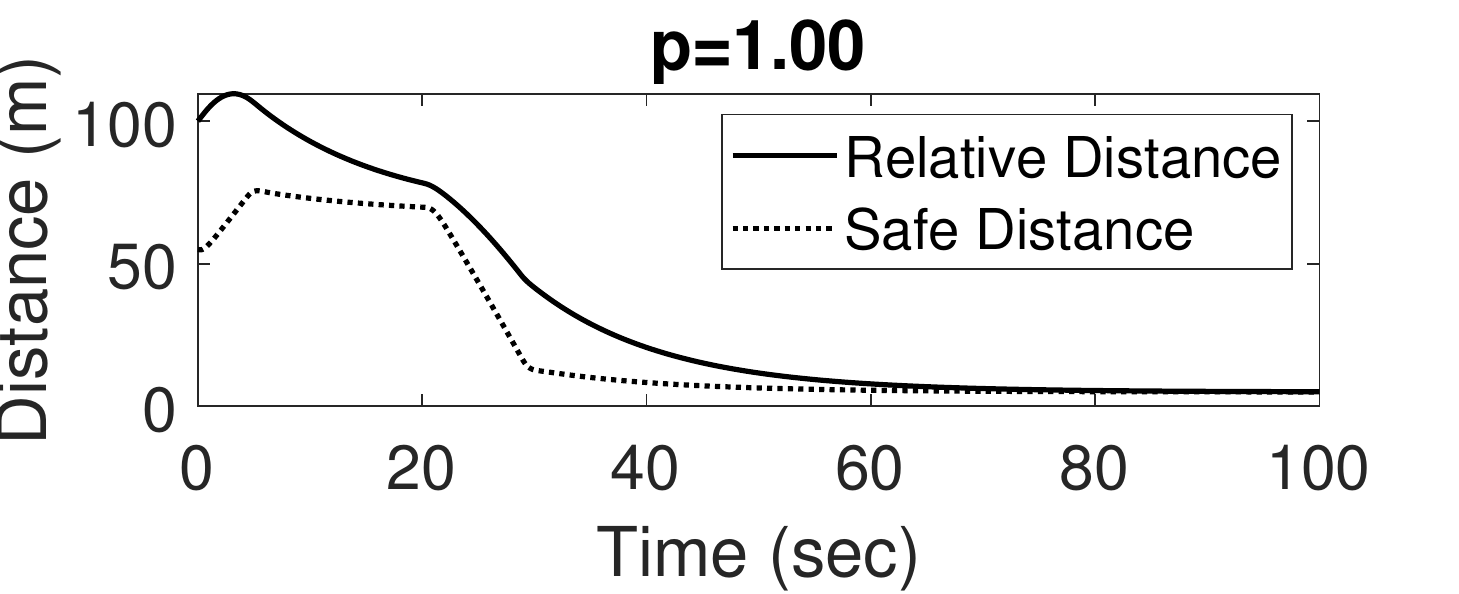}
    \caption{\label{fig_p1} Distance Keeping Performance (Reference).}
\end{figure}

The driving scenario tested is whether the ACC function can successfully stops the vehicle when the lead car performs an emergency brake. The initial position is set to be $100m$ for the lead car, and $0m$ for the ego car. The initial velocity is set to be $25m/s$ (56mph) for the lead car, and $20m/s$ (45mph) for the ego car. The lead car will maintain a constant speed for the first 20 seconds and then perform a brake with constant acceleration at $-2.5m/s^2$ until its velocity reaches zero. In general, we expect the ego car in this scenario to accelerate at the beginning to catch up the lead car while keeping a safe distance according to $d_\text{des}$. After the lead car starts to brakes, we also expect the ego car to decelerates and stops at a safe stopping distance from the lead car, which is $d_0=5m$. Without the attacker in the system and the controller periodically transmits control signals ($p=1$), the performance is shown in Figure \ref{fig_p1}. Based on this figure, we observe that the relative distance gradually converges to the safe distance for the first 20 seconds and when the lead vehicle starts to brake, the ego car successfully maintained a safe distance and stops 5m behind the lead car at the end of the simulation. When the attacker is present, we will then use the performance shown in Figure \ref{fig_p1} as the reference model to compare with.

\begin{figure}[h]
    \centering
    \includegraphics[width=0.4\textwidth]{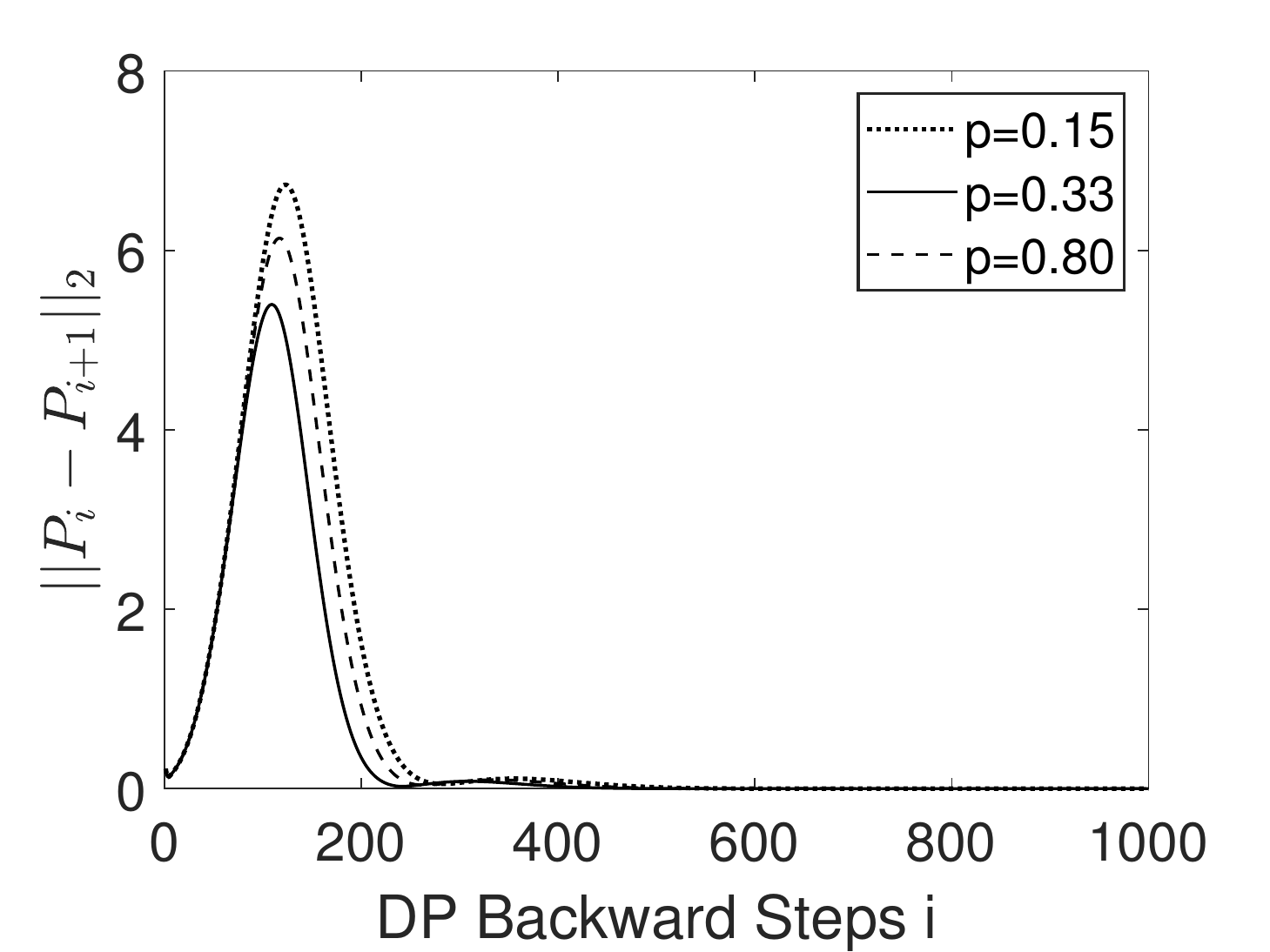}
    \caption{ \label{fig_Pconvergence} Convergence of Value Function (1 step = $100ms$).}
\end{figure}

\begin{figure}[!t]
     \centering
     \begin{subfigure}[b]{0.4\textwidth}
         \centering
         \includegraphics[width=\textwidth]{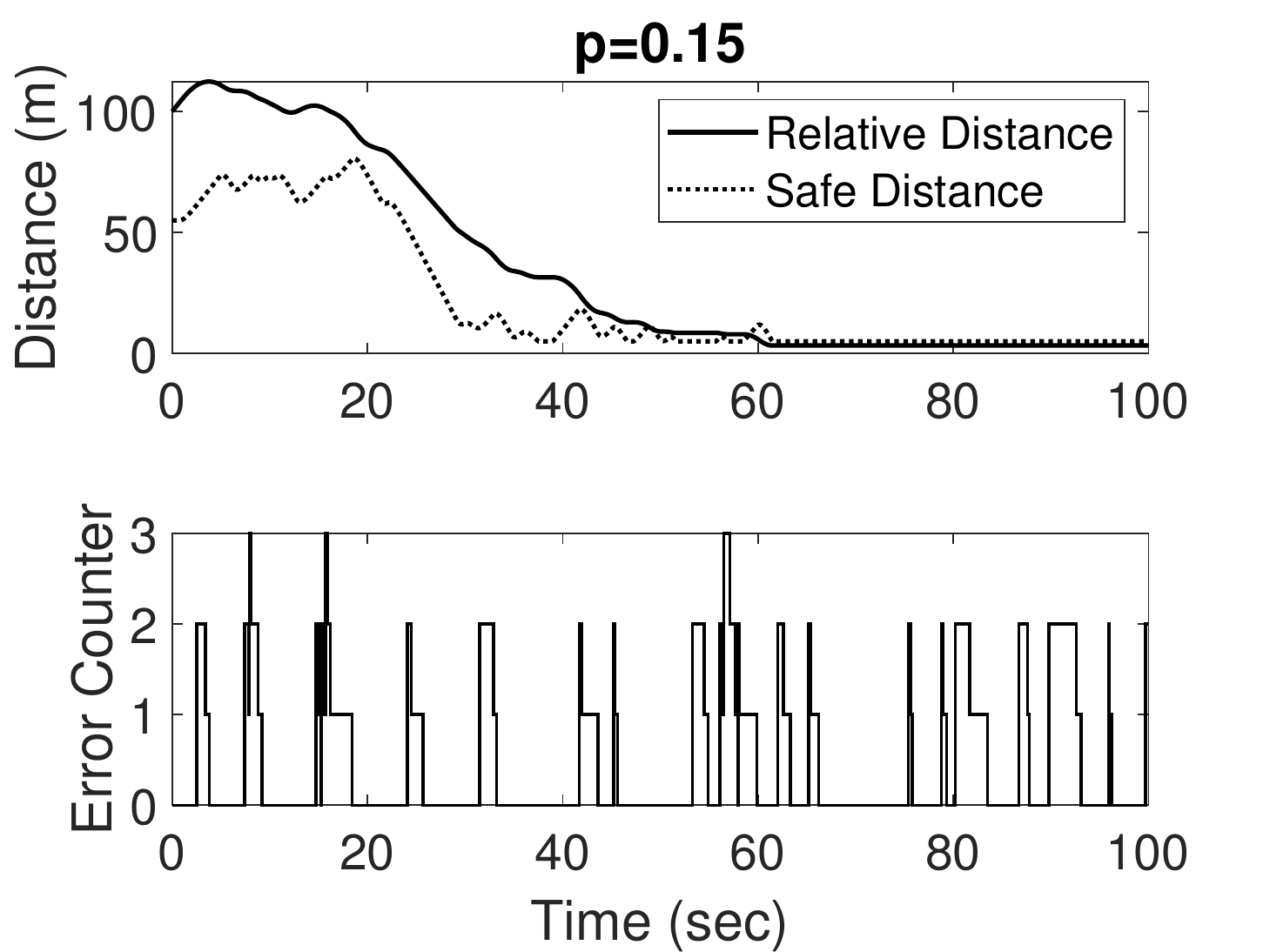}
         \caption{}
         \label{fig_p015}
     \end{subfigure}
     \\
     \begin{subfigure}[b]{0.4\textwidth}
         \centering
         \includegraphics[width=\textwidth]{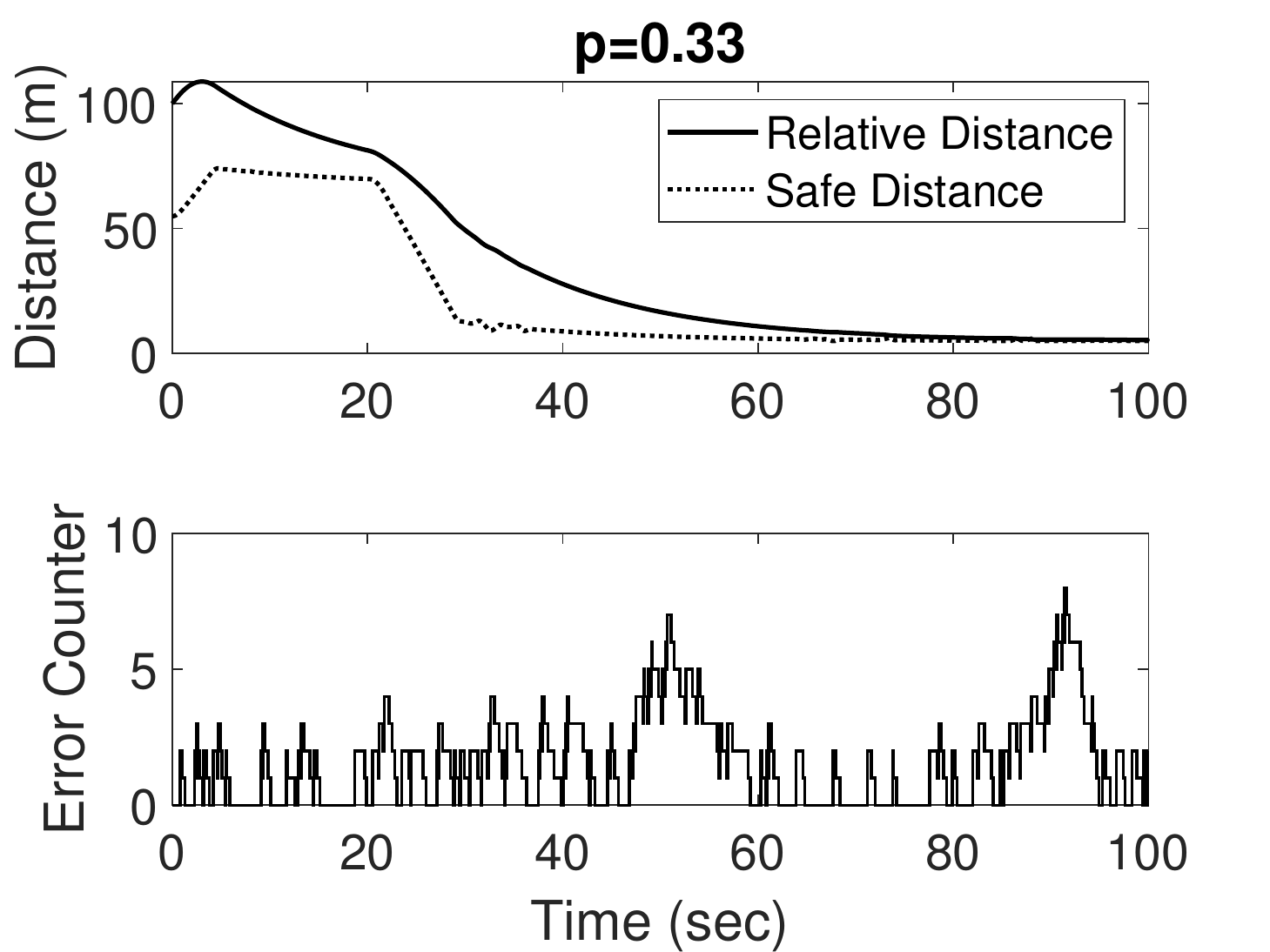}
         \caption{}
         \label{fig_p033}
     \end{subfigure}
     \\
     \begin{subfigure}[b]{0.4\textwidth}
         \centering
         \includegraphics[width=\textwidth]{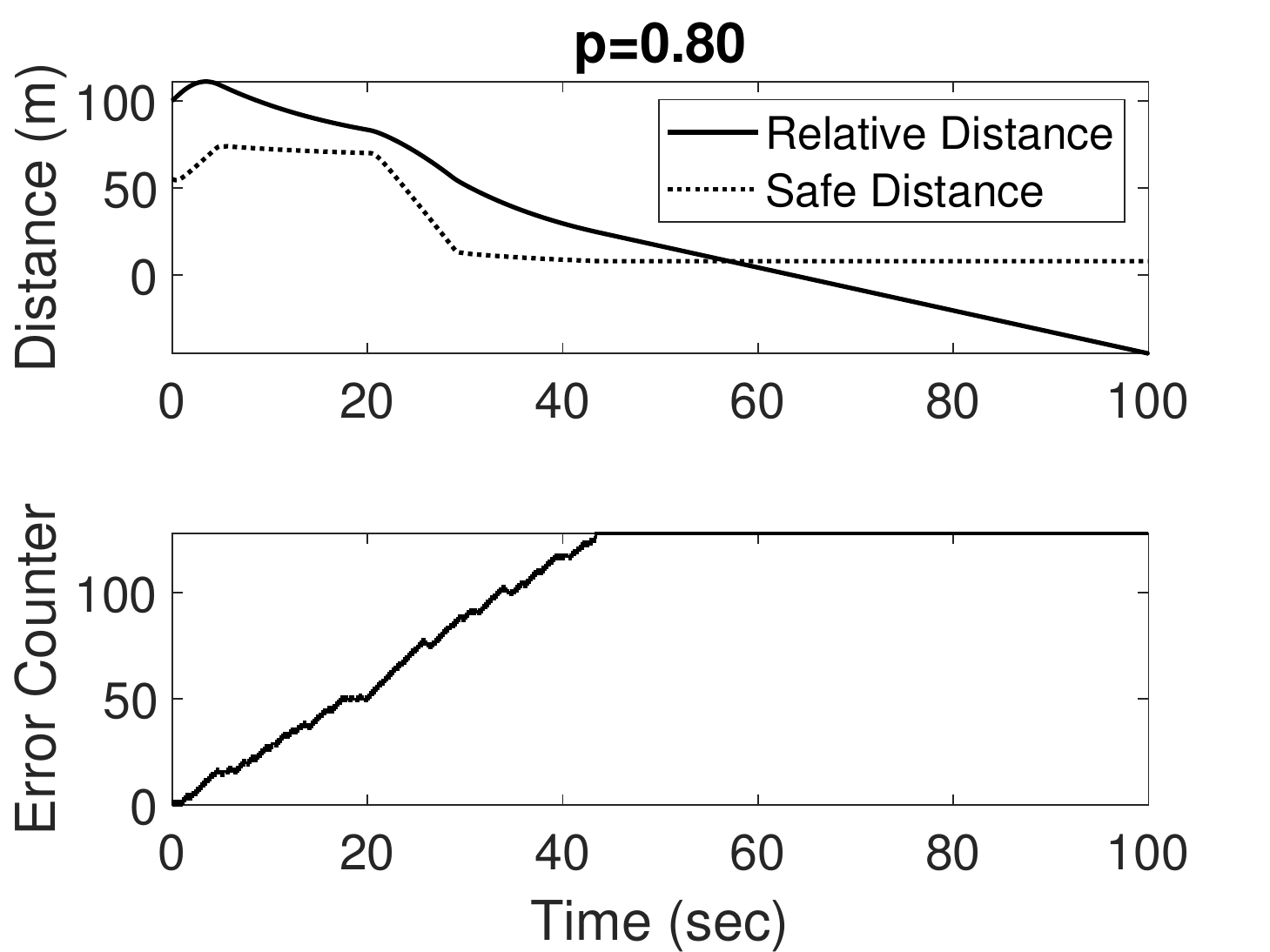}
         \caption{}
         \label{fig_p08}
     \end{subfigure}
        \caption{Distance Keeping and Error Performance Using Stochastic Transmission}
        \label{fig_simulation}
\end{figure}

Now consider there is a closed-loop attacker present in the system, which follows the dominant attack strategy discussed in Section \ref{sec_dominantAttackProof}. The control and attack policy follows the Nash equilibrium, and the resulting dynamics is described in \eqref{eqn_statespaceWattack}. Recall the error counter formulated in Section \ref{sec_MCerrorCounter}, we set $e_+=2$ as the penalty score when the attacker successfully triggers a collision and on the other hand $e_-=-1$ when the transmitter of the controller successfully delivered a message. The threshold $\bar{e}$ to trigger a bus-off event is set to be $128$. The attacker is assumed to be in the system through the whole simulation.

Next, we will talk about how to find a reasonable range of $p$ for the transmission policy of the controller. In general, the probability of transmitting control signals cannot be too small such that it is too intermittent to stabilize the system. On the other hand, such probability cannot be too large such that the attacker gains enough collisions in the network to trigger the bus-off attack. 

We first use the necessary condition found in \eqref{eqn_pBound} as a lower bound of picking $p$ for the transmitter. This can be interpreted as the range of the probability of transmitting a message for the controller at each time step that is necessary for a bounded average cost. It turns out that $\lambda_{\max}(A)=1$ which implies $p(1-p)>0$ is the necessary condition for bounded average cost so there is no constraint of picking $p$ in this case according to \eqref{eqn_pBound}. However, it is worth noting that in some applications, it is possible to have the lower bound larger than 0.25 for $p(1-p)$ which leads to no possible values of $p\in(0,1)$ that satisfies the necessary condition for bounded average cost. In these cases, one potential solution is to decrease the sampling time such that the $A$ matrix of the linear system is closer to the identity matrix, which has a $\lambda_{\max}(A)$ closer to 1.

An upper bound on $p$ can also be derived if we restrict a negative drift of the error counter based on the values of $e_+$ and $e_-$. This can be written as $pe_++(1-p)e_-<0$, or in this case $p<1/3$. One remark here is that a negative drift of the error counter does not guarantee the system is free of bus-off event. In fact, as long as the error counter is bounded below by zero, bus-off event will eventually happen (with probability 1) given any fixed positive threshold if the attacker stays in the system as $t\rightarrow\infty$. The upper bound is chosen such that the bus-off attack occurs with sufficiently small probability in a finite horizon. Now with these suggestions, we pick $p=0.15, 0.33, 0.8$ as the three choices for $p$. As discussed later, $p=0.15$ ($p=0.8$) can be considered as overly conservative (optimistic) transmission policy against bus-off attacker. Given the three choices of $p$, the convergence of the value functions in Lemma \ref{thm_optimalFinite} is numerically checked. As shown in Figure \ref{fig_Pconvergence}, this is done by calculating the 2-norm of the error between two consecutive $P_k$ matrices. This implies that all the three values of $p$ picked above yield to a bounded cost.

The ACC performance with an emergency brake of the lead car is then simulated using $p=0.15, 0.33, 0.8$. As shown in Figure \ref{fig_simulation}, the performance is measured by the relative distance and the value of error counter within 100 seconds, or equivalently 1000 simulation steps with 10hz sampling frequency. For $p=0.33$ (Figure \ref{fig_p033}), we see that the error counter is below 10, and the bus-off attack is avoided. The relative distance is also kept reasonably close to the safe distance compared with the reference model shown in Figure \ref{fig_p1}. For $p=0.15$ (Figure \ref{fig_p015}), the error counter is also very low due to an even smaller value of $p$ picked. However, the ego car stopped about $3.3m$ behind the lead car which is below the desired stopping distance which is $d_0=5m$. For $p=0.8$ (Figure \ref{fig_p08}), due to a positive drift of the error counter, a bus-off event happened at $t=43.4s$ and the control signal is lost afterward. As a result, the acceleration is out of control and the vehicle crashed in to the lead car around $t=63.6s$, which is considered as the first time relative distance is below 0. 

\section{Conclusion} \label{sec_conclusion}
In this paper, we introduced stochastic transmission as a defense scheme against bus-off attack in CAN networks. Simulation results shows that using an appropriate transmission probability, the error counter can be maintained at a low level without triggering the bus-off event. Further, under certain assumptions on the unstable eigenvalues of the system and the transmission probability, the system can be made stable using the subgame-perfect Nash equilibrium control policy.

\begin{appendices}
    \section{Proof of Dominant Attack Strategy} \label{app_dominantAttack}
    To show $\gamma_A^*$ as the unique minimizer, we first fix $p$ and define the probability of collision $q(\gamma_A)$ as
    \begin{equation*}
        q(\gamma_A)\triangleq\mathbb{P}(t_i^A=t_i^\tx|t_i^\tx\sim\text{Geometric}(p),t_i^A\sim\gamma^A).
    \end{equation*}
    The first part of the proof will show that $\gamma_A^*$ defined in Equation \eqref{eq:gammaA} is the unique maximizer of $q(\gamma_A)$ for any fixed $p$. Notice that $q(\gamma_A)$ can be derived as
    \begin{align*}
        q(\gamma_A)=&\mathbb{P}(t_i^A=t_i^\tx)=\sum_{k=1}^\infty\mathbb{P}(t_i^A=t_i^\tx=k)\\
        \overset{(a)}{=}&\sum_{k=1}^\infty\mathbb{P}(t_i^A=k)\mathbb{P}(t_i^\tx=k)\overset{(b)}{=}\sum_{k=1}^\infty\iota_k p(1-p)^{k-1}\\\overset{(c)}{\leq}&\sum_{k=1}^\infty\iota_kp=p,
    \end{align*}
    where the equality (a) holds since $t_i^A$ and $t_i^\tx$ are independent according to the restriction of controller's and attacker's transmission policy. The equality (b) holds due to $t_i^\tx$ are geometrically distributed. Now we notice that $\{p(1-p)^{k-1}\}_k$ is a decreasing sequence of $k\geq 1$, and $\sum_{k=1}^\infty \iota_k=1$, then $q$ is maximized if and only if $\iota_1=1$. In addition, the inequality (c) is tight, then we have $\max q = p$. The above discussion implies for any $p\in(0,1)$, we have
    \begin{equation*}
        \begin{aligned} 
            \gamma_A^*=&\arg\max_{\gamma_A\in\Gamma_A} q(\gamma_A)=\{\mathbb{P}(t_i^A=1)=1, \forall i\geq1\},\\
            p=&\max_{\gamma_A\in\Gamma_A} q(\gamma_A).
        \end{aligned}
    \end{equation*}
    where $\gamma_A^*$ is the unique maximizer. 
    
    Next, we will show that if the attacker wants to minimize $J_A(\gamma_C,\gamma_A)$, then it is equivalent to maximize $q(p,\gamma_A)$, which leads to the dominant attack strategy as $\gamma_A^*$. In the remaining proof, we will use $q$ instead of $q(p,\gamma_A)$ as the probability of collision for simplicity.

    Based on Section \ref{sec_MCerrorCounter}, the transition probability matrix of the error counter $\{S_i\}_i$ as a Markov process is given by $\Theta(q)=[\theta_{ss'}(q)]_{s,s'\in\mathcal{S}}$, where
    \begin{align*}
        \theta_{ss'}(q)=
        \begin{cases}
        q & \text{ if } s'=s+e_+\\
        1-q & \text{ if } s'=s+e_-\text{ or } s=s'=0\\
        1 & \text{ if } s=s'=\bar e\\
        0 & \text{ otherwise }
        \end{cases}.
    \end{align*}
    
    Note that $\bar e$ is an absorbing state and all the other states are transient. We can then assign a reward $1$ for each transition from $s\in\mathcal{S}\setminus\{\bar e\}$ to $s'\in\mathcal{S}$ and $0$ reward to the transition from $s\in\mathcal{S}$. In this case, starting from any state $s_0\in\mathcal{S}\setminus\{\bar e\}$, the expected accumulated rewards in the steady state of the Markov chain equals to the expected first hitting time to the state $\bar e$. Denote the accumulated rewards vector as $\boldsymbol{v}(q)=[v_s(q)]_{s\in\mathcal{S}}$, where $v_s(q)$ is the accumulated rewards starting from state $s$. Then in the steady state, we have
    \begin{align}\label{eq:reward}
        \boldsymbol{v}(q)=\boldsymbol{1}+\Theta(q) \boldsymbol{v}(q).
    \end{align}
    In this case, the expected steps of bus-off event conditioned on the probability of collision is $\mathbb{E}[\xi|q]=v_0(q)$.
    
    Next we will show that equation \eqref{eq:reward} has a unique solution and it is monotonically decreasing with respect to $q$. Since $\bar e$ is an absorbing state, we can then transform $\Theta(q)$ into the following Jordan canonical form:
    \begin{align*}
        \Theta(q)=\left[
        \begin{array}{ccc|c}
        & & & \\
        & \bar \Theta(q) & & \boldsymbol{\Tilde{{1}}}_{\bar{k}e_+} \\
        & & & \\\hline
        & \boldsymbol{0} & & 1
        \end{array}\right],
    \end{align*}
    where $\bar k$ is such that $\bar ke_+\leq \bar e<(\bar k+1)e_+$, and $\bar \Theta$ is the transition probability of all the transient states $\mathcal{S}\setminus\{\bar e\}$. Here $I-\bar \theta$ is invertible according to \cite[Theorem 11.4, p418]{grinstead2012introduction}. That is, equation \eqref{eq:reward} can be simplified by removing the absorbing state $\bar e$, which is
    \begin{align*}
        \boldsymbol{v}(q)_{\mathcal{S}\setminus\{\bar e\}}=\boldsymbol{1}+\bar \Theta(q) \boldsymbol{v}(q)_{\mathcal{S}\setminus\{\bar e\}}=\left(I-\bar\Theta(q)\right)^{-1}\boldsymbol{1}.
    \end{align*}
    
    According to \cite{Miller1981}, let $g$ be such that
    \begin{align*}
        g=&\trace{\left(-\bar\Theta(q)\right)I^{-1}}=-\trace{\bar \Theta(q)}=q-1,
    \end{align*}
    then
    \begin{align*}
    \left(I-\Theta(q)\right)^{-1}=&\left(I^{-1}-\frac{1}{1+g}I^{-1}\left(-\bar\Theta(q)\right)I^{-1}\right)\\
    =&I+\frac{1}{q}\bar\Theta(q).
    \end{align*}
    Thus, we have
    \begin{align*}
        v_0(q)=1+\frac{1}{q}\sum_{i=0,j\in\mathcal{S}\setminus\{\bar e\}}\theta_{ij}=1+\frac{1}{q},
    \end{align*}
    which shows that $v_0(q)$ is monotonically decreasing with respect to $q$. Therefore, minimizing $\mathbb{E}(\xi|)$ is equivalent to maximizing $q(\gamma_A)$, and $\gamma_A^*$ is the unique maximizer of $q(\gamma_A)$. That is, for any $q\in(0,1)$,
    \begin{align*}
        \ex{\xi|\gamma_A^*}=v_0(p)\leq v_0(q)=\ex{\xi|\gamma_A},
    \end{align*}
    which proves the result.
\end{appendices}

\bibliographystyle{ieeetr}
\bibliography{main.bib}

\begin{IEEEbiography}[{\includegraphics[width=1in,height=1.25in,clip,keepaspectratio]{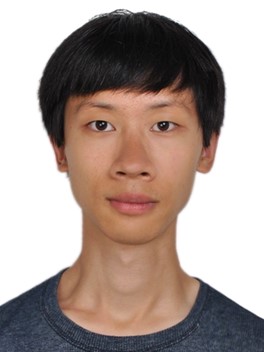}}]{Jiacheng Tang}
received the B.S. degree in Applied Mathematics, the B.S. and M.Sc. degree in Electrical and Computer Engineering, all from The Ohio State University, Columbus Ohio, in 2016, 2016, and 2017 respectively. Since 2017, he has been with The Ohio State University, where he is currently a Ph.D. student in Electrical and Computer Engineering under supervision of Prof. Abhishek Gupta. His research interests are in the area of cyber security for control system, optimization algorithms, and statistical inference.
\end{IEEEbiography}

\begin{IEEEbiography}[{\includegraphics[width=1in,height=1.25in,clip,keepaspectratio]{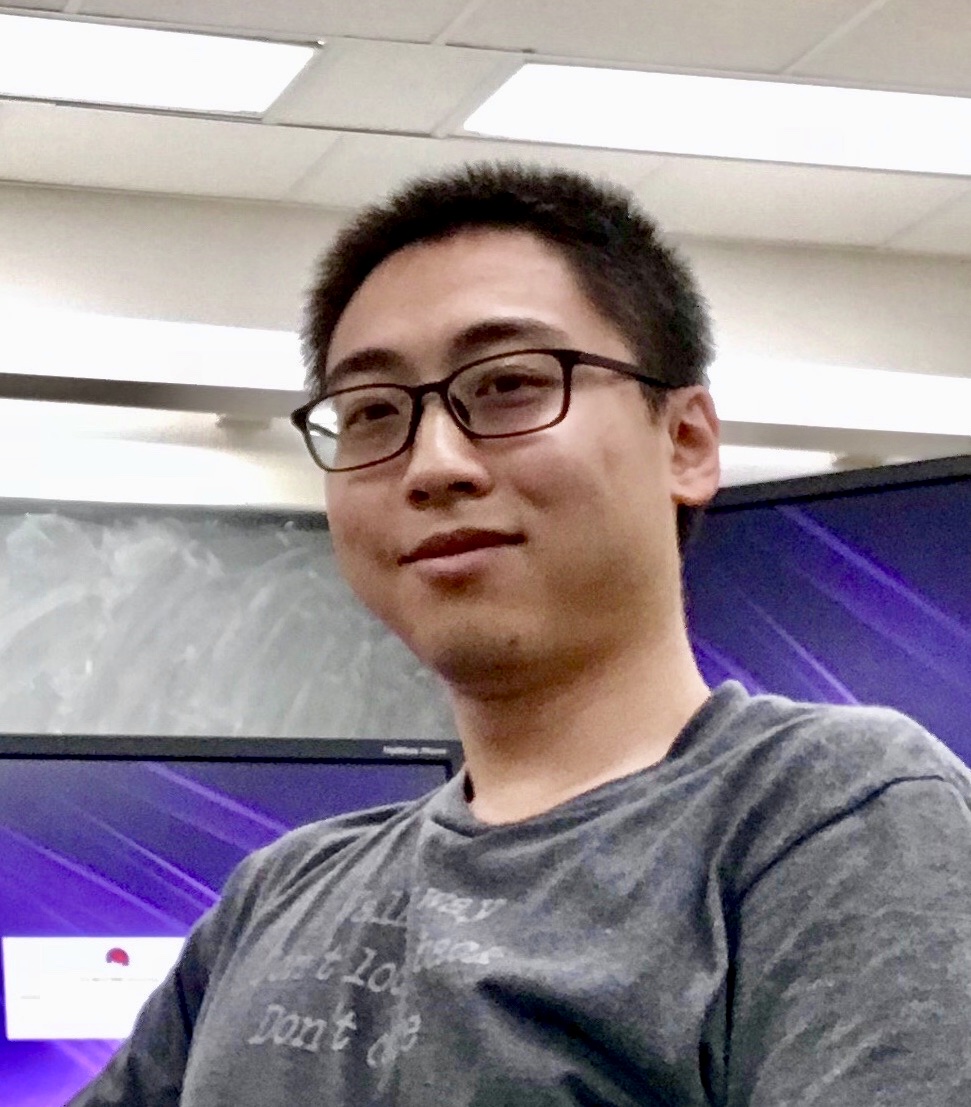}}]{Shiping Shao}
received his B.S. degree and M.S. degree in Nanchang University in 2014 and 2016, respectively. Currently, he is a Ph.D. student in the department of Electrical and Computer Engineering at The Ohio State University and under supervision of Prof. Abhishek Gupta. His research interests are in the area of optimization algorithms and stochastic control systems with application in market design.
\end{IEEEbiography}

\begin{IEEEbiography}[{\includegraphics[width=1in,height=1.25in,clip,keepaspectratio]{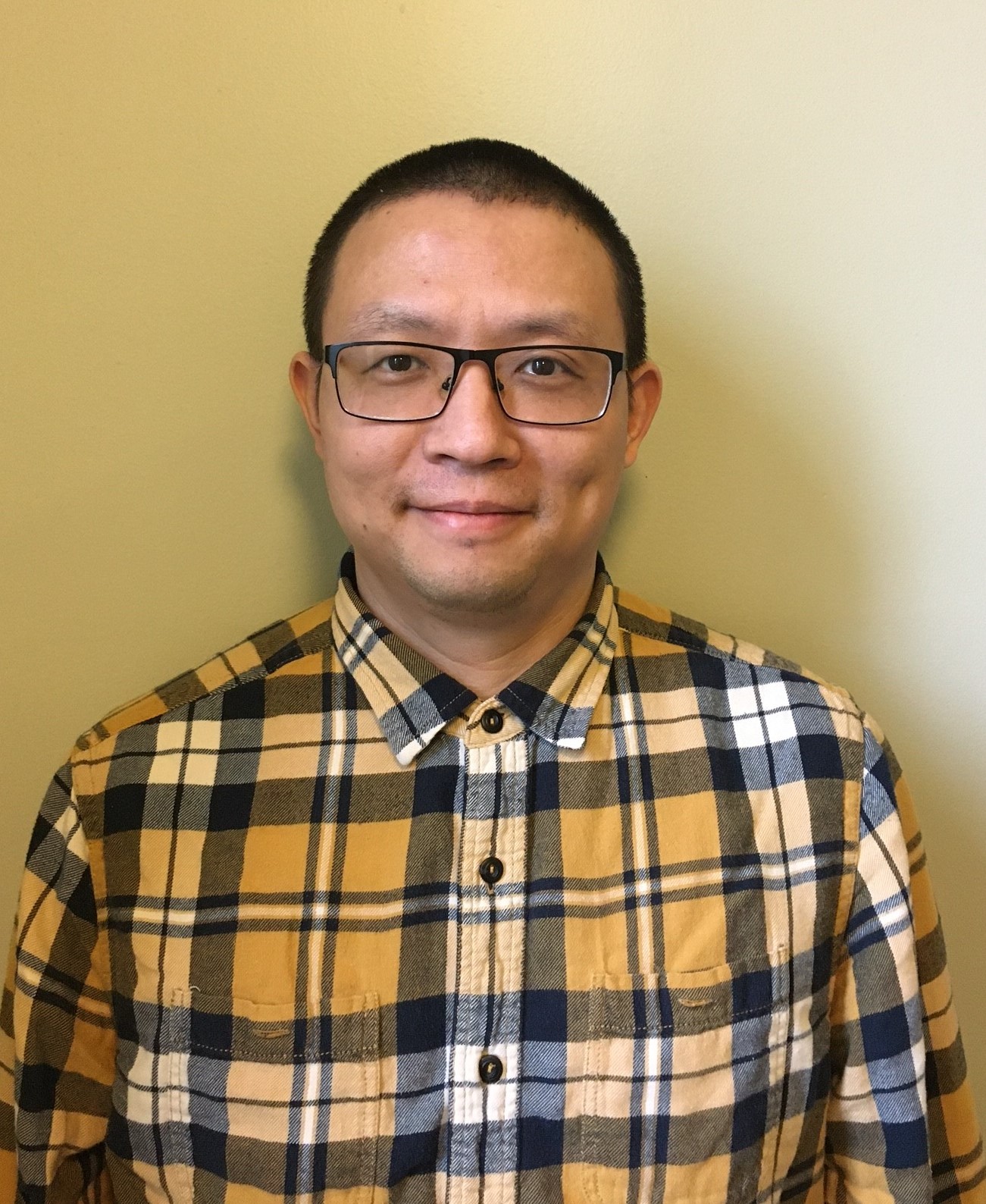}}]{Jiguo Song} received his PhD degree in Computer Science for his work on System-level Fault Tolerance for Real-time Operating System (RTOS) from George Washington University in 2016. He joined Ford Motor Company Research\&Advanced Engineering Department in 2017 as a security research engineer, and currently work with Ford In-vehicle Core Software Architecture team. His work at Ford has focused on automotive system dependability, including CAN-network Intrusion Detection System and In-vehicle Software Control-flow Protection.
\end{IEEEbiography}

\begin{IEEEbiography}[{\includegraphics[width=1in,height=1.25in,clip,keepaspectratio]{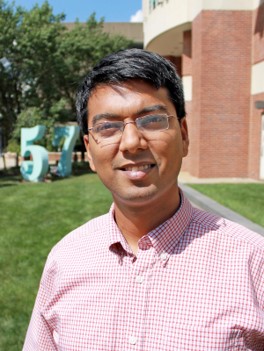}}]{Abhishek Gupta}
Abhishek Gupta is an assistant professor in the ECE department at The Ohio State University. He completed his MS and PhD in Aerospace Engineering from University of Illinois at Urbana-Champaign (UIUC) in 2014, MS in Applied Mathematics from UIUC in 2012, and B.Tech. in Aerospace Engineering from IIT Bombay in 2009. His research interests are in stochastic control theory, probability theory, and game theory with applications to transportation markets, electricity markets, and cybersecurity of control systems.
\end{IEEEbiography}

\end{document}